\documentclass[12pt]{article}
\usepackage{amssymb}

\usepackage[dvips]{epsfig}
\usepackage[dvips]{graphics}
\newcommand{\text}{\rm}

\begin{document}

\title{\textbf{Discussing the Spectrum of the Kalb-Ramond Field Coupled to
3D-Gravity}}
\author{\textbf{J.L. Boldo$^{a}$}\thanks{\texttt{E-mail: jboldo@cce.ufes.br}}\textbf{%
, J.A. Helay\"{e}l-Neto$^{b,c}$}\thanks{\texttt{E-mail: helayel@gft.ucp.br}}%
\textbf{\ and N. Panza}$^{c}$\vspace{2mm} \\
\textbf{$^{a}$}UFES, Universidade Federal do Esp\'{\i }rito Santo\\
CCE, Departamento de F\'{\i}sica\\
Campus Universit\'{a}rio de Goiabeiras 29060-900\\
Vit\'{o}ria, ES, Brazil\vspace{2mm}\\
\textbf{$^{b}$}CBPF, Centro Brasileiro\textbf{\ }de Pesquisas F\'{\i }sicas 
\\
Rua Xavier Sigaud 150, 22290-180 Urca \\
Rio de Janeiro, Brazil\vspace{2mm}\\
\textbf{$^{c}$}UCP, Universidade Cat\'{o}lica de Petr\'{o}polis\\
Rua Bar\~{a}o do Amazonas 124, 25685-070\\
Petr\'{o}polis, Brazil\vspace{2mm}}
\maketitle

\begin{abstract}
The mechanism of dynamical mass generation for the gauge field is studied
through 1-loop. We find out that torsion is an obstruction to the appearance
of a 1-loop mass correction. Contrary, if torsion is not present, a mass gap
is generated for the 2-form field.

\setcounter{page}{0}\thispagestyle{empty}
\end{abstract}

\vfill\newpage\ \makeatother
\renewcommand{\theequation}{\thesection.\arabic{equation}} %
\renewcommand{\baselinestretch}{2}

\section{\ Introduction\-}

The Abelian 2-form gauge potential, usually referred to as Kalb-Ramond field
(K-R), exhibits a number of interesting features in four dimensional
space-time (4D) \cite{K-R}; special emphasis is given to the fact that it
mixes up with the electromagnetic gauge potential so as to yield a massive
spin-1 excitation based on a $\left[ U\left( 1\right) \right] ^2$-symmetry,
without the need for Higgs scalars \cite{Cremmer}. More recently, the
coupling between the K-R and Maxwell fields has been reconsidered in
connection with the issue of Dirac-like monopoles with massive photons \cite
{WA}. Also, interesting discussions on the possible non-Abelian extension of
the K-R field are set in the works of Ref. \cite{Sorella}.

Reassessing the K-R field in three dimensional space-time (3D) brings about
some peculiarities that show up as a by-product of three dimensional
space-time. For example, the Abelian gauge field dual to the K-R potential
in 3D is subject to a particular gauge symmetry that selects its
longitudinal component as the physical propagating mode, whereas its
transverse part appears as a compensating mode, and so it can always be
gauged away. Once the mass-shell condition is imposed, no physical
propagating degree of freedom survives and we have the peculiarity that the
K-R gauge field carries \textit{no} on-shell degrees of freedom in $3$D \cite
{Torsion2}. We understand that this is a consequence of setting its dynamics
on a flat space-time. So, our propose is to set out this paper in order to
illustrate how the coupling of the K-R field to the 3D counterpart of the
Einstein-Hilbert gravity excites a massive scalar mode in the spectrum of
the theory. This is a result that already emerges at classical level.

The main motivation of our work is to compute 1-loop corrections to the K-R
field self-energy, so as to understand how these effects change the mass
spectrum set up at tree-level. We pursue our investigation in the cases of
ordinary (torsionless) gravity (Section 2) and, later on, we go over to the
general case of gravity with torsion (Section 3) \cite{Kerlick}. We find
some peculiarities at the end of our analysis and we shall comment on them
in our Conclusive Comments (Section 4).

\section{The Riemannian Case}

The model we contemplate accounts for the non-minimal coupling of $3$%
D-gravity to the K-R field according to the Lagrangian density given below: 
\begin{equation}
\mathcal{L}=\mathcal{L}_{EH}+\mathcal{L}_{gauge}+\mathcal{L}_{int}\,,
\label{2.1}
\end{equation}
where the first term is the usual Einstein-Hilbert Lagrangian: 
\begin{equation}
\mathcal{L}_{EH}=\frac 1{\kappa ^2}\sqrt{g}R{.}  \label{2.2}
\end{equation}
Notice the absence of the overall minus sign that appears in $4$D-gravity:
in $3$D, this must be the choice in such a way to avoid that the graviton
becomes a ghost.

The second term describes the Lagrangian for the antisymmetric gauge field $%
B_{\mu \nu }$, 
\begin{equation}
\mathcal{L}_{gauge}=\frac{1}{6}\sqrt{g}G_{\mu \nu \lambda }G^{\mu \nu
\lambda },  \label{2.3}
\end{equation}
where $G_{\mu \nu \lambda }$ is a $3$-form written in terms of the potential 
$B_{\mu \nu }$ as follows: 
\begin{equation}
G_{\mu \nu \lambda }=\nabla _{\mu }B_{\nu \lambda }+\nabla _{\lambda }B_{\mu
\nu }+\nabla _{\nu }B_{\lambda \mu }\,.  \label{2.4}
\end{equation}
The last term accounts for the interaction between gravity and the gauge
field: 
\begin{equation}
\mathcal{L}_{int}=\xi \sqrt{g}\epsilon ^{\mu \nu \lambda }\nabla _{\mu
}B_{\nu \lambda }R,  \label{2.5}
\end{equation}
with $\epsilon ^{\mu \nu \lambda }=\frac{\varepsilon ^{\mu \nu \lambda }}{%
\sqrt{g}}$, $\varepsilon ^{\mu \nu \lambda }$ being the totally
antisymmetric symbol, while $\nabla _{\mu }$ denotes the covariant
derivative. The coupling constant $\xi $ carries dimension of (mass)$^{-%
\frac{1}{2}}$.

To analyse the espectrum of the model, we consider the linearised
approximation of the full theory. This consists in the expansion of the
metric field around the flat background as below: 
\begin{equation}
g_{\mu \nu }=\eta _{\mu \nu }+\kappa h_{\mu \nu }\mathrm{\ },  \label{2.6}
\end{equation}
where $h_{\mu \nu }$ is the perturbation associated to the graviton field.

As for the $2$-form gauge potential, it is more convenient to parametrise it
in terms of its dual, $B_\mu =\frac 1{3!}\epsilon _{\mu \nu \lambda }B^{\nu
\lambda }$. Doing so, the gauge transformation for $B_{\mu \nu }$ can be
rephrased as 
\begin{equation}
\delta B_\lambda =\epsilon _\lambda {}^{\mu \nu }\nabla _\mu \chi _\nu 
\mathrm{\ },  \label{2.7}
\end{equation}
where $\chi _\mu $ is the $3$-vector gauge parameter.

In order to invert the wave operator that mixes $h_{\mu \nu }$ and $B_\mu $
in the bilinear piece of the action, we fix the De Donder gauge for $h_{\mu
\nu },$

\begin{equation}
\mathcal{L}_{Donder}=\frac 1{2\alpha }F_\mu F^\mu \mathrm{\ },  \label{2.9}
\end{equation}
where 
\begin{equation}
F_\mu =\partial _\lambda \left( h_\mu ^\lambda -\frac 12\delta _\mu ^\lambda
h\right) .  \label{2.10}
\end{equation}

On the other hand, the gauge transformations ($\ref{2.7}$) for $B_\mu $
suggests the following gauge-fixing term for the Abelian symmetry associated
to $B_{\mu \nu }$: 
\begin{equation}
\mathcal{L}_{gf}=\beta \left( \epsilon ^{\mu \nu \lambda }\nabla _\nu
B_\lambda \right) ^2.  \label{2.17}
\end{equation}

With these elements, the bilinear piece of the action can be cast under the
form: 
\begin{equation}
\mathcal{L}=\frac{1}{2}\sum_{\alpha \beta }\phi ^{\alpha }\mathcal{O}%
_{\alpha \beta }\phi ^{\beta },  \label{13}
\end{equation}
where $\phi =\left( h^{\mu \nu },B^{\mu }\right) $ and $\mathcal{O}$ is the
differential wave operator to be inverted to give us the graviton and the
gauge field propagators. This operator can be decomposed in four sectors,
namely, 
\begin{equation}
\mathcal{O}=\left( 
\begin{array}{cc}
A_{\mu \nu ,\kappa \lambda } & B_{\mu \nu ,\kappa } \\ 
C_{\mu ,\kappa \lambda } & D_{\mu ,\nu }
\end{array}
\right) ,  \label{2.23}
\end{equation}
where 
\begin{eqnarray*}
A_{\mu \nu ,\kappa \lambda } &=&\frac{\square }{2}P_{\mu \nu ,\kappa \lambda
}^{\left( 2\right) }-\frac{\square }{2\alpha }P_{\mu \nu ,\kappa \lambda
}^{\left( 1\right) }-\frac{\left( \alpha +1\right) \square }{2\alpha }P_{\mu
\nu ,\kappa \lambda }^{\left( 0-s\right) }-\frac{\square }{4\alpha }P_{\mu
\nu ,\kappa \lambda }^{\left( 0-w\right) } \\
&&+\frac{\sqrt{2}\square }{4\alpha }\left( P_{\mu \nu ,\kappa \lambda
}^{\left( 0-sw\right) }+P_{\mu \nu ,\kappa \lambda }^{\left( 0-ws\right)
}\right) ,
\end{eqnarray*}
\[
B_{\mu \nu ,\kappa }=-2\kappa \xi \square \theta _{\mu \nu }\partial
_{\kappa }, 
\]
\begin{equation}
C_{\mu ,\kappa \lambda }=2\kappa \xi \square \partial _{\mu }\theta _{\kappa
\lambda },  \label{2.24}
\end{equation}
\[
D_{\mu ,\nu }=-2\square \left( \beta \theta _{\mu \nu }+\omega _{\mu \nu
}\right) , 
\]
where the $P^{\prime }$s are the Barnes-Rivers projector operators for
symmetric second-rank tensors in D=3 (see Ref. \cite{Torsion1} for details),
while $\theta _{\mu \nu }$ and $\omega _{\mu \nu }$ are respectively the
transverse and longitudinal operators for vectors.

The propagators are given by 
\begin{equation}
\left\langle 0\left| T\left[ \Phi _{\alpha }\left( x\right) \Phi _{\beta
}\left( y\right) \right] \right| 0\right\rangle =i\left( \mathcal{O}%
^{-1}\right) _{\alpha \beta }\delta ^{3}\left( x-y\right) ,  \label{2.25}
\end{equation}
where wave operator is inverted by using of the multiplicative relations
satisfied by the projectors \cite{Torsion1}. In so doing, the propagators in
momentum space read as follows: 
\begin{eqnarray*}
\left\langle hh\right\rangle &=&\frac{2i}{p^{2}}\left\{ -P^{\left( 2\right)
}+\alpha P^{\left( 1\right) }+\frac{2\left[ p^{2}-m^{2}\left( \alpha
+1\right) \right] }{\left( p^{2}-m^{2}\right) }P^{\left( 0-w\right) }\right.
\\
&&-\left. \frac{m^{2}}{\left( p^{2}-m^{2}\right) }\left[ P^{\left(
0-s\right) }+\sqrt{2}\left( P^{\left( 0-sw\right) }+P^{\left( 0-ws\right)
}\right) \right] \right\} ,
\end{eqnarray*}

\begin{equation}
\left\langle h_{\mu \nu }B_\kappa \right\rangle =\frac 1{4\kappa \xi
p^2\left( p^2-m^2\right) }\left( \theta _{\mu \nu }+2\omega _{\mu \nu
}\right) p_\kappa ,  \label{2.28}
\end{equation}

\[
\left\langle B_\mu B_\nu \right\rangle =\frac i{2\beta p^2}\theta _{\mu \nu
}-\frac{im^2}{2p^2\left( p^2-m^2\right) }\omega _{\mu \nu }, 
\]
where $m^2=\frac 1{8\kappa ^2\xi ^2}$ and we have suppressed the indices in $%
h_{\mu \nu }$ and in the $P_{\mu \nu ,\kappa \lambda }$ projectors.

The above propagators display poles at $p^2=0$ and $p^2=m^2>0$, so that
tachyons do not show up. The next step is a investigation of the possibility
that negative-norm state might be present.

A necessary criterium for unitarity at tree-level is analysed by saturating
the propagator with external currents $J^{\mu \nu }$ and $J^\mu $ (for the
graviton and the gauge field, respectively), compatible with the gauge
symmetries of the Lagrangian. Ghosts are absent in the model if the
imaginary part of the residue of the current-current transition amplitude
taken at the propagator poles is non-negative. For the graviton, this
amplitude in momentum space is written as 
\begin{equation}
\mathcal{A}\equiv J^{*\mu \nu }\left( p\right) \left\langle T\left[ h_{\mu
\nu }\left( -p\right) h_{\kappa \lambda }\left( p\right) \right]
\right\rangle J^{\kappa \lambda }\left( p\right) ,  \label{2.29}
\end{equation}
where, by virtue of the transversality of $J^{\mu \nu }\left( p\right) $,
only the spin projectors $P^{\left( 2\right) }$ and $P^{\left( 0-s\right) }$
survive. This fact enable us to write the imaginary part of the residue of
the transition amplitude at the pole $p^2=0$ as 
\begin{equation}
I\,m(\mathcal{R}es\mathcal{A})=2\left( \left| J^{\mu \nu }\right| ^2-\left|
J_\nu ^\nu \right| ^2\right) {.}  \label{2.30}
\end{equation}

Now, defining the following set of independent vectors in momentum space, 
\begin{equation}
p^\mu \equiv \left( p^0,\mathbf{p}\right) ,\mathrm{\ }\overline{p}^\mu
\equiv \left( p_0,-\mathbf{p}\right) ,\ \epsilon ^\mu \equiv \left( 0,%
\mathbf{\epsilon }\right) ,\mathrm{\ }  \label{2.31}
\end{equation}
that satisfy the conditions: 
\begin{eqnarray}
\bar{p}.p &=&\left( p^0\right) ^2+\left( \mathbf{p}\right) ^2>0\mathrm{\ },
\label{2.32} \\
p.\epsilon &=&\bar{p}.\epsilon =0,  \nonumber \\
\epsilon ^\mu \epsilon _\mu &=&-1,  \nonumber
\end{eqnarray}
the symmetric tensor current $J^{\mu \nu }\left( p\right) \,$can be
decomposed according to: 
\begin{eqnarray}
J_{\mu \nu }\left( p\right) &=&a\left( p\right) p_\mu p_\nu +b\left(
p\right) p_{(\mu }\bar{p}_{\nu )}+c\left( p\right) p_{(\mu }\epsilon _{\nu
)}+  \nonumber \\
&&d\left( p\right) \bar{p}_\mu \bar{p}_\nu +e\left( p\right) \bar{p}_{(\mu
}\epsilon _{\nu )}+f\left( p\right) \epsilon _{(\mu }\epsilon _{\nu )}.
\label{2.33}
\end{eqnarray}

Substituting (\ref{2.33}) into (\ref{2.30}), and making use the relations (%
\ref{2.31}) and (\ref{2.32}), we find that $I\,m(\mathcal{R}es\mathcal{A})=0$%
, showing that the massless pole does not propagate. For the pole at $%
p^2=m^2 $, one obtains 
\begin{equation}
I\,m(\mathcal{R}es\mathcal{A})=\left| J_\mu ^\mu \right| ^2\mathrm{\ },
\label{2.34}
\end{equation}
that is always positive-definite. From this result, we find only one
on-shell degree of freedom for the massive graviton.

For the vector field, the transition amplitude is given by 
\begin{equation}
\mathcal{A}=J^{*\mu }\left( p\right) \left\langle T\left[ B_\mu \left(
-p\right) B_\nu \left( p\right) \right] \right\rangle J^\nu \left( p\right) .
\label{2.35}
\end{equation}
Expanding the current $J^\mu \left( p\right) $ with respect to the basis (%
\ref{2.31}), 
\begin{equation}
J^\mu \left( p\right) =a\left( p\right) p^\mu +b\left( p\right) \overline{p}%
^\mu +c\left( p\right) \varepsilon ^\mu \ ,  \label{2.36}
\end{equation}
and by making use of its conservation law, 
\begin{equation}
\epsilon ^{\mu \nu \lambda }p_\nu J_\lambda =0\mathrm{\ },  \label{2.37}
\end{equation}
we can show that the massless pole is again non-dynamical, while the residue
of the amplitude at the massive pole give us: 
\begin{equation}
\frac 12\left| a\right| ^2m^2\mathrm{\ },  \label{2.39}
\end{equation}
which ensures one physical degree of freedom. Therefore, the non-minimal
coupling of the K-R field to gravity results in a appearance of a dynamical
massive pole in the longitudinal sector of $B_{\mu \nu }$.

We now turn into the calculation of the 1-loop mass correction for the gauge
field propagator. The relevant vertices to our discussion are shown in
figure 1, while all possible self-energy corrections are depicted in figure
2. The vertices of the figures 1.a and 1.b come from the Lagrangians (\ref
{2.3}) and (\ref{2.5}) by expanding the metric up to order $\kappa $, while
the one of figure 1.c is obtained from (\ref{2.3}) up to order $\kappa ^2$.

\hspace{-2cm}\scalebox{1}{\includegraphics[86,580][300,720]{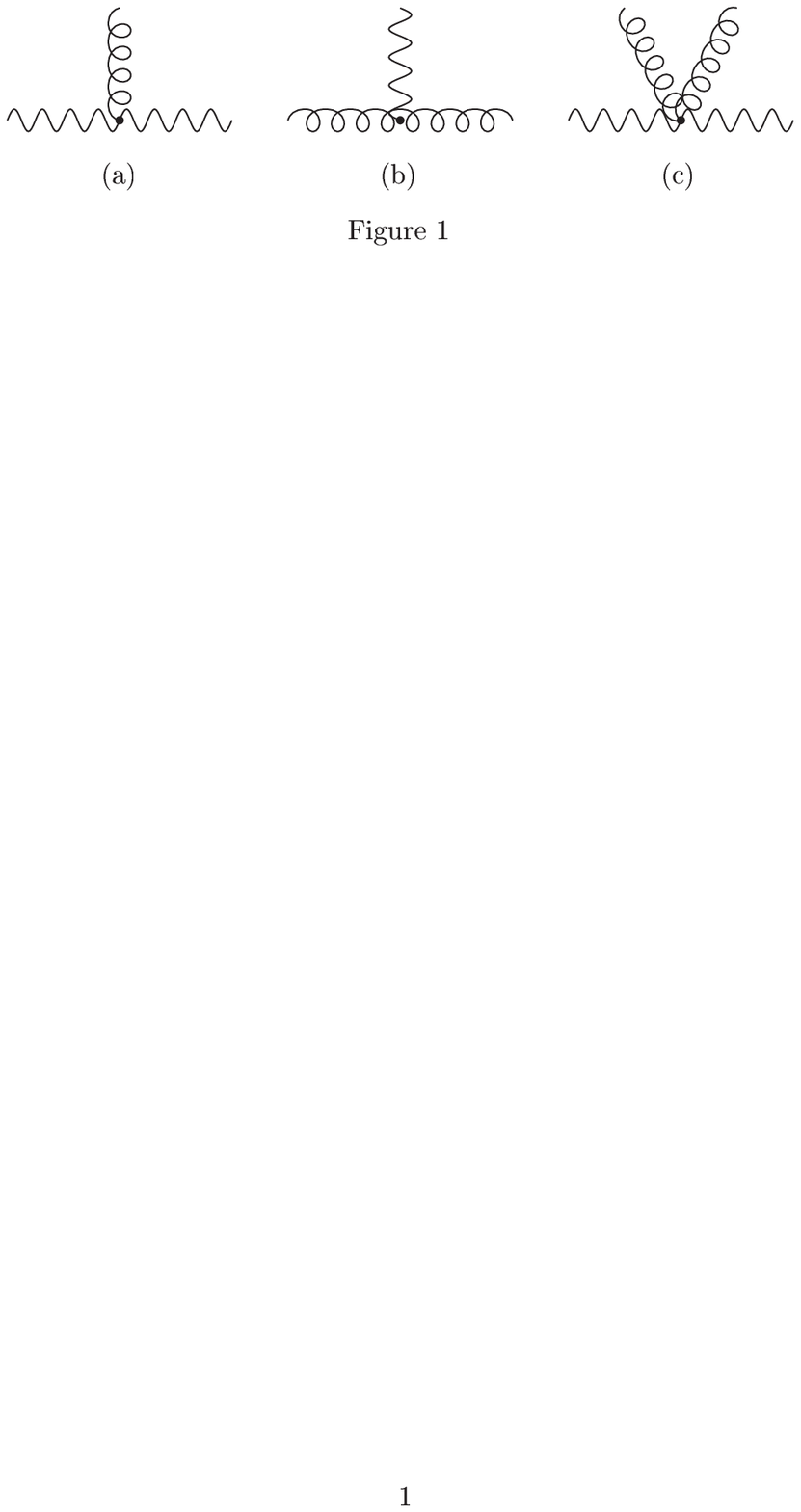}}

\hspace{-1cm}\scalebox{1}{\includegraphics[86,620][300,740]{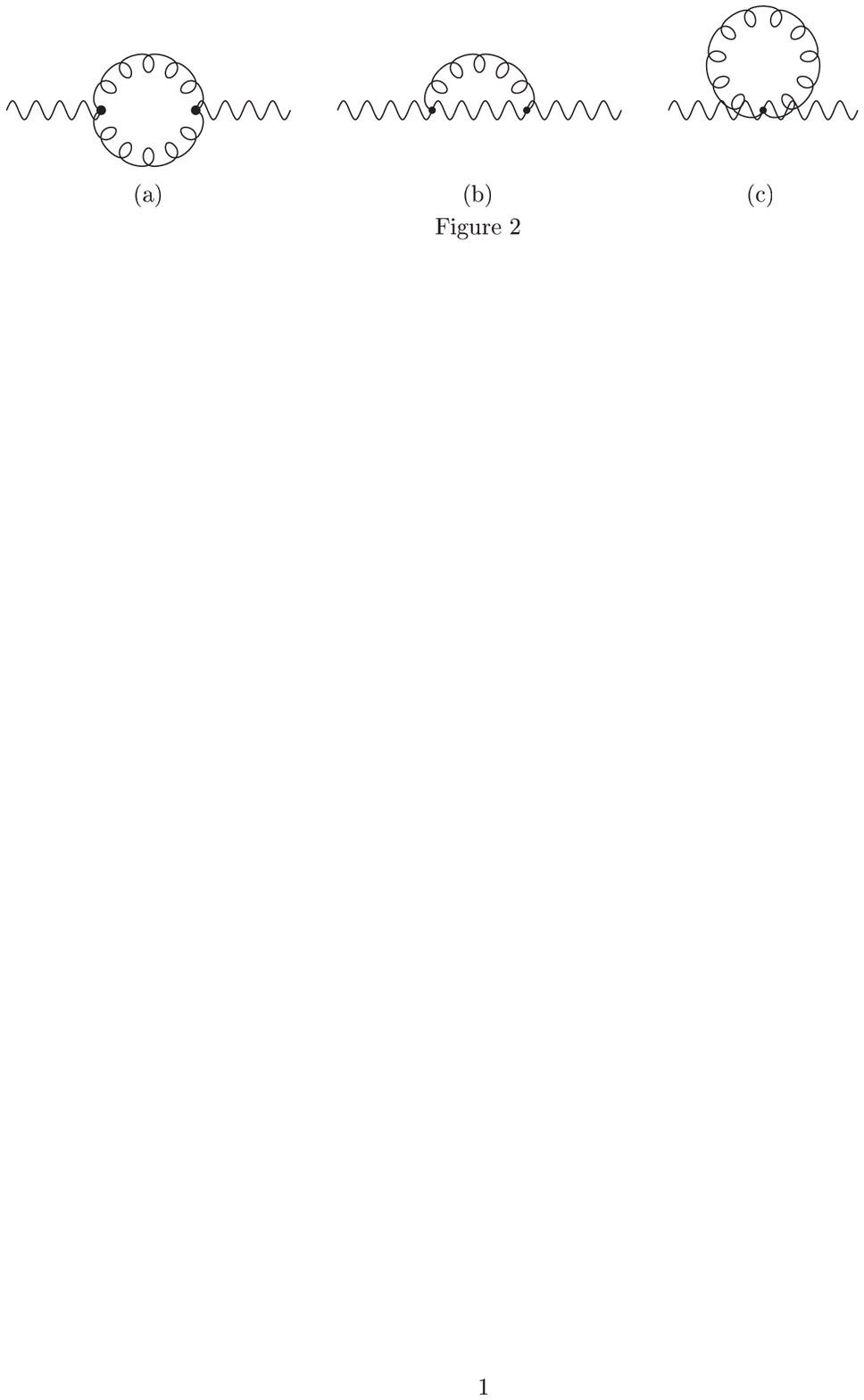}}

From the Feynman rules displayed in figure 1, the contribution coming from
the loop integrals can be evaluated. Due to algebraic complexity of the
vertices and propagators, the explicit evaluation of these diagrams only was
possible by means of the software FORM. Since we are concerned with the
contribution to the vector field mass, and we checked that a
Chern-Simons-like term is \textit{not} generated, we can calculate the
diagrams by setting all momenta appearing on the external legs to zero. One
therefore, obtains the following contribution to the 1-loop shift in the
gauge field mass: 
\begin{equation}
\left. \Delta \,m^2\right| _{1-loop}=\frac{3\pi ^2}{16\sqrt{2}\kappa \xi ^3}.
\label{2.43}
\end{equation}

\section{The Non-Riemannian Case}

Now, following the same procedure as the one in the previous section, we
propose to extend our analysis to the case where torsion is non-vanishing.
In three-dimensional space-time, the torsion field can be covariantly split
according to its SO(1,2) irreducible components: 
\begin{equation}
T_{\mu \nu \lambda }=\varepsilon _{\mu \nu \lambda }\varphi +\frac 12\left(
\eta _{\nu \lambda }t_\mu -\eta _{\mu \lambda }t_\nu \right) +\varepsilon
_{\mu \nu \alpha }X^\alpha {}_\lambda \mathrm{\ },  \label{3.1}
\end{equation}
namely, a trace part $t_\mu =T_{\mu \nu }$ $^\nu $, a totally antisymmetric
part $\varphi =\frac 1{3!}\varepsilon ^{\mu \nu \lambda }T_{\mu \nu \lambda
} $ and a traceless rank-2 symmetric tensor $X_{\mu \nu }$ \cite{Torsion1}.

A possible gauge-invariant action may be so chosen that there is no coupling
at all between the K-R field and torsion, i.e., the former ``neither yields
nor feels torsion''. Thus, let us consider the action 
\begin{equation}
\mathcal{S}=\int d^3x\sqrt{g}\left( \frac 1{\kappa ^2}R+\frac 16G_{\mu \nu
\lambda }G^{\mu \nu \lambda }+2\xi R\widetilde{\nabla }_\mu B^\mu \right) ,
\label{3.2}
\end{equation}
where the tilde on $\nabla _\mu $ means that we are considering only the
Riemaniann part of the covariant derivative. Varying the total action (\ref
{3.2}) with respect to the independent fields, we get the following result
for the trace part of the torsion: 
\begin{equation}
t^\mu =-\frac{4\kappa ^2\xi \widetilde{\nabla }^\mu \widetilde{\nabla }_\nu
B^\nu }{1+2\kappa ^2\xi \widetilde{\nabla }_\nu B^\nu },  \label{3.3}
\end{equation}
while the variations with respect to the fields $\varphi $ and $X_{\mu \nu }$
give us 
\begin{equation}
\left\{ 
\begin{array}{c}
\varphi =0\mathrm{\ }, \\ 
X_{\mu \nu }=0\mathrm{\ }.
\end{array}
\right.  \label{3.4}
\end{equation}
The torsion irreducible components act as mere auxiliary fields and one can
eliminate them by means of their respective algebraic field equations.
Therefore, substituing this results back into the action, one obtain a
Lagrangian expressed exclusively in terms of $B_\mu $ and $h_{\mu \nu }$: 
\begin{eqnarray}
\mathcal{L} &=&\frac 1{\kappa ^2}\sqrt{g}\widetilde{R}+\frac 16\sqrt{g}%
G_{\mu \nu \lambda }G^{\mu \nu \lambda }+2\xi \sqrt{g}\widetilde{R}%
\widetilde{\nabla }_\mu B^\mu  \label{3.5} \\
&&+8\kappa ^2\xi ^2\sqrt{g}\frac{\widetilde{\nabla }_\lambda \widetilde{%
\nabla }_\mu B^\mu \widetilde{\nabla }^\lambda \widetilde{\nabla }_\nu B^\nu 
}{1+2\kappa ^2\xi \widetilde{\nabla }_\mu B^\mu }\mathrm{\ }.  \nonumber
\end{eqnarray}

From the bilinear sector of the Lagrangian above supplemented by the gauge
fixing terms (\ref{2.9}) and (\ref{2.17}), we obtain the following
propagators: 
\begin{eqnarray}
\left\langle hh\right\rangle &=&\frac{2i}{p^2}\left\{ -P^{\left( 2\right)
}+\alpha P^{\left( 1\right) }+2\left[ \left( \alpha +1\right) +\frac{p^2}{m^2%
}\right] P^{\left( 0-w\right) }\right.  \nonumber \\
&&+\left. \left( 1+\frac{p^2}{m^2}\right) \left[ P^{\left( 0-s\right) }+%
\sqrt{2}\left( P^{\left( 0-sw\right) }+P^{\left( 0-ws\right) }\right)
\right] \right\} ,  \label{3.6}
\end{eqnarray}

\[
\left\langle B_\mu B_\nu \right\rangle =\frac i{2p^2}\left( \frac 1\beta
\theta _{\mu \nu }+\omega _{\mu \nu }\right) . 
\]

It is remarkable to notice now that the poles are located at $p^2=0$,
contrary to the previous case, where torsion was not present. It is just the
last term of the action (\ref{3.5}) the responsible for the suppression of
the massive pole. The mixing between $h_{\mu \nu }$ and $B_\mu $ arising
from the mentioned term appears with the right coefficient that eliminates
the massive pole.

Proceeding analogously to what we did in the previous Section, the massless
poles are shown to be non-dynamical for both fields. From this result, one
concludes that the coupling of the K-R field to gravity with non-vanishing
torsion may result in a theory without on-shell degrees of freedom at
tree-level as long as the action (\ref{3.5}) is concerned. Moreover, by
computing the Feynman rules for this action and calculating the 1-loop
self-energy graphs, we checked that no mass is dynamically generated by
radiative corrections.

\section{Concluding Remarks}

As a final result, we can state that the coupling of the 3D K-R field with
Einstein-Hilbert gravity excites a dynamical mode of the rank-2 gauge
potential, whenever torsion is not considered. The inclusion of torsion as a
non-dynamical field changes this result, as we checked in Section 3.
However, had we considered non-minimal couplings with higher powers in the
curvature, the situation would be completely changed, as torsion would
become dynamical and its mixing with the graviton \cite{Torsion1} would for
sure trigger a dynamical mass for the gauge field on the basis of the
results worked out in Section 2.

To conclude, we would like to mention that, if gravity were described by a
pure Chern-Simons term, even in the torsionless case the K-R field would
remain non-dynamical on-shell, since no mass would be generated neither at
the classical nor at the 1-loop level.

\end{document}